\documentclass[aps,prd,twocolumn,superscriptaddress,longbibliography,nofootinbib]{revtex4-2}
\usepackage[utf8]{inputenc}
\usepackage{color}
\usepackage{graphicx}
\usepackage{subfigure}
\usepackage{xcolor}
\usepackage[normalem]{ulem}
\usepackage[pdftex,breaklinks,colorlinks,linkcolor=blue,citecolor=teal,anchorcolor=red,urlcolor=cyan]{hyperref}
\usepackage{orcidlink}
\usepackage{multirow}

\usepackage{amsmath,amssymb,amsfonts}

\def\prl{Phys. Rev. Lett.}
\def\prd{Phys. Rev. D}

\def\apj{Astrophys. J.}

\def\apjl{Astrophys. J. Lett.}

\def\pau_p{Prog. Theor. Phys.}

\def\mnras{Mon. Not. R. Astron. Soc.}

\def\physrep{Phys. Rep.}
\def\nat{Nature}

\def\jcap{J. Cosmology Astropart. Phys}

\def\nar{New Astronomy Reviews}
\def\araa{Annual Review of Astronomy and Astrophysics}
\def\aj{Astronomical J.}

\def\mchirp{\mathcal M}
\begin{document}

\title{Subsolar mass black holes from stellar collapse induced by primordial black holes}
\author{Thomas W.~Baumgarte\orcidlink{0000-0002-6316-602X}}
\email{tbaumgar@bowdoin.edu}
\affiliation{Department of Physics and Astronomy, Bowdoin College, Brunswick, Maine 04011, USA}

\author{Stuart L.~Shapiro\orcidlink{0000-0002-3263-7386}}
\email{slshapir@illinois.edu}
\affiliation{Department of Physics, University of Illinois at Urbana-Champaign, Urbana, Illinois 61801}
\affiliation{Department of Astronomy and NCSA, University of Illinois at Urbana-Champaign, Urbana, Illinois 61801}

\begin{abstract}
While no gravitational-wave detection of subsolar mass black holes has been confirmed to date, a number of candidate detections invite us to speculate on the origin of such black holes should a detection be confirmed.  It is generally assumed that the observation of a black hole with subsolar mass $M_{\rm obs}$ would provide strong evidence for primordial black holes (PBHs).  The mass $M_{\rm PBH}$ of the PBH, however, does not necessarily have to be equal to $M_{\rm obs}$, as it would in what we term a  ``direct PBH scenario". Instead, a black hole of mass $M_{\rm obs}$ may form in a capture of a much smaller primordial black hole, $M_{\rm PBH} \ll M_{\rm obs}$, by a dwarf star of mass $M_* \simeq M_{\rm obs}$, followed by the total consumption of the star by the PBH. We provide some rough estimates and demonstrate that such an ``indirect PBH scenario" may also lead to significant populations of black holes with mass $M_{\rm obs}$, especially in dwarf galaxies, and may be able to explain rare subsolar mass events.
\end{abstract}

\maketitle

Starting with the first direct detection of gravitational waves in 2015 \cite{LIGO_GW150914}, the LIGO-Virgo-Kagra (LVK) Collaboration has observed numerous gravitational wave signals.  Most of these signals were emitted by binaries of stellar-mass black holes (with masses greater than about $5 M_\odot$), and some by either neutron star or mixed binaries.  

The LVK Collaboration has also conducted searches for subsolar mass objects.  Given that the minimum mass of neutron stars is expected to be around $1.1 M_\odot$ (see, e.g., \cite{SuwYSUT18} and references therein, but note the unusually small mass reported for the neutron star at the center of the supernova remnant HESS J1731-347, e.g., \cite{DorSPS22}), the detection of a compact object with mass well below a solar mass could very well be a black hole.  In addition to arguments based on the minimum mass of neutron stars, the presence or absence of tidal effects could also be used to distinguish subsolar mass black holes from neutron stars (see \cite{Cre25}).  Because small-mass binaries merge at high frequencies, at which current gravitational-wave detectors are less sensitive, the detection and analysis of their gravitational wave signals may require special instruments and techniques (see \cite{Aggetal25,MilP25}).

Since subsolar mass black holes are not expected to form in supernova explosions, a confirmed detection of subsolar mass black holes would point to a primordial origin of the black hole.  In fact, as discussed in \cite{CarCGHK24}, a detection of a subsolar mass black hole, or a black hole at very high redshift, would provide direct evidence for the existence of primordial black holes (PBHs).

\begin{table*}
    \begin{tabular}{c|c|c|c|c|c|c}
         Event & $m_1~[M_\odot]$ & $m_2~[M_\odot]$ & $\mchirp~[M_\odot]$ & FAR [yr$^{-1}$] & SNR & Reference\\
         \hline
         SSM170401 & 4.71 & 0.76 & [1.42, 1.57] & 0.41 & 8.67 & \cite{Phuetal21,Moretal23} \\
         SSM200308 & 0.62 & 0.27 & [0.351, 0.355] & 0.20 & 8.90 & 
         \cite{Pruetal24} \\
         S251112cm &  & & [0.1, 0.87] & 0.16 &  & \cite{2025GCN.42650....1L,GraceDB} 
    \end{tabular}
    \caption{Selected candidates for subsolar mass binaries.   In addition to the event name we list, where available, estimates of the primary's and secondary's mass $m_1$ and $m_2$, the chirp mass $\mchirp$, the false alarm rate FAR, the signal-to-noise ratio SNR, as well as some relevant references. See, e.g., \cite{NitW21,Phuetal21,LVK23,CarCGHK24} for additional candidates with larger FARs.  }
    \label{tab:candidates}
\end{table*}

Motivated by these considerations, several authors have analyzed and re-analyzed gravitational wave data searching for low-mass binaries -- in the words of \cite{Phuetal21}, ``the hunt for subsolar primordial black holes \ldots~is open".  To date, no observation of potential gravitational wave signals from low-mass binaries has a false alarm rate (FAR) that is small enough for the signal to be considered a detection with sufficient confidence (see, e.g., \cite{LIGO19,LVK23,NitW21,NitW22}).  In the meantime, however, several candidate signals have been identified, albeit with large FARs.  In Table \ref{tab:candidates} we list three of these candidates with (relatively) small FARs.  Two of these, SSM170401 and SSM200308 have been discussed in the literature before (e.g.~\cite{Phuetal21,Moretal23,Pruetal24,YuaH24}), while S251112cm, the signal with the smallest FAR, is a very recent detection that, to the best of our knowledge, has not been fully analyzed yet.  We also refer to \cite{Phuetal21,NitW21,CarCGHK24,LVK23} for more potential subsolar mass binary signals.  

%The primary in SSM170401, observed in LIGO's second observing run (O2), is probably a stellar-mass black hole, but the companion has a mass below the minimum mass for neutron stars discussed above

While we have yet to record any gravitational wave signal with a sufficiently small FAR to be confirmed as originating from subsolar mass black holes with sufficient confidence, we are invited to speculate on the possible origin of such a black hole, should such a signal be confirmed in the future.  As discussed above, such a detection would strongly suggest a primordial origin of the black hole.

Strong constraints on the contribution of primordial black holes to the dark matter content of the Universe exist in many mass ranges, but other masses remain either unconstrained or only weakly constrained.  Adopting the notation in Fig.~1 of \cite{CarK20} (see also \cite{CarKSY21}), PBHs could make up the entire dark matter in window A, $10^{-16} M_\odot \lesssim M_{\rm PBH} \lesssim 10^{-10} M_\odot$.   There are two other mass windows, window B around $M_{\rm PBH} \simeq 10^{-6} M_\odot$ (i.e.~Earth-mass black holes) and window C around $M_{\rm PBH} \simeq 10^{-1} M_\odot$, in which PBHs could make up up to about 10\% of the dark matter.  Assuming a detection of a black hole with mass $M_{\rm obs} \simeq 10^{-1} M_\odot$, for example, it is then tempting to conclude that this black hole belongs to a population of PBHs formed in window C, so that $M_{\rm obs} \simeq M_{\rm PBH}$.  This ``direct PBH scenario" arising in the very early universe is adopted by most authors who consider the primordial nature of subsolar mass black holes, although, according to \cite{YuaH24}, such a PBH population may also lead to conflicts with pulsar-timing array observations.  

In this short note we describe an alternative ``indirect PBH scenario", in which a black hole of mass $M_{\rm obs}$ is formed from the collision of a much smaller PBH, so that $M_{\rm PBH} \ll M_{\rm obs}$, with a star of mass $M_* \simeq M_{\rm obs} \simeq 10^{-1} M_\odot$, i.e.~a dwarf star.  The capture of such a PBH would ultimately lead to the consumption of the star by the PBH, forming a black hole of mass $M_{\rm obs}$ (see, e.g., \cite{AbrBBGJQ09,OncMGG22,BauS25c} and references therein). Because PBHs with small masses could be quite numerous, and because dwarf stars are both numerous and have long lifetimes, this process could also lead to a significant number of black holes with masses around $10^{-1} M_\odot$, and hence could potentially also explain subsolar mass black holes -- should their existence be confirmed.

The exact rates of PBH collisions with and capture by stars depend sensitively on the astrophysical environment, in particular the dispersion velocities of the stars and the dark matter.  In order to illustrate the indirect PBH scenario we will consider dwarf galaxies, which have small dispersion velocities and high dark-matter content (see, e.g., \cite{McC12,Sim19} for reviews of dwarf and ultra-faint dwarf galaxies), both of which are conducive to PBH collision and capture by stars.  We will focus on collisions between PBH and main-sequence stars here, but note that PBH capture during the (far shorter) star-formation process provides another avenue for forming sub-solar mass black holes, and may allow the capture of PBHs with masses smaller than the ones we consider here (see, e.g., \cite{EssT23}).

A typical value for the stellar dispersion velocity in dwarf galaxies is about $\sigma \simeq 10\,\mbox{km/s}$ (e.g.~\cite{WalMOGWSW07,BatHB13}), even though some are also significantly smaller (see also Table~1 in \cite{Sim19}).  For the sample of dwarf galaxies studied by \cite{Stretal08}, the dynamical (i.e.~total) mass within 300 pc from their centers was found to be around $10^7\, M_\odot$, resulting in a dark-matter density of about $\rho_{\rm DM} \simeq 0.1 \, M_\odot/\mbox{pc}^3 \simeq 7 \times 10^{-24}\,\mbox{g/cm}^3$, even though some models predict a higher central dark-matter density (see Fig.~2 in \cite{WalMOGWSW07}).  The total (baryonic) mass in stars varies significantly from dwarf galaxy to dwarf galaxy. For Fornax, for example, it is estimated to be a few time $10^7$ (see, e.g., Table 4 in \cite{McC12} and Table 1 in \cite{ReaWS19}), from which we estimate a typical number of stars of around $N_* \simeq 5 \times 10^7$.  Finally, identifying the total mass of a dwarf galaxy with its virial mass, a typical number is around $M_{\rm tot} \simeq 10^9 M_\odot$ (see, e.g., the numbers compiled in Table 1 of \cite{WalMOGWSW07}).  We provide some specific data for a selection of dwarf galaxies in Table \ref{tab:galaxies} below.

% For the Fornax dwarf galaxy orbiting the Milky Way, for example, the dispersion velocity is about $v_\infty \simeq 10\,\mbox{km/s}$ (e.g.~\cite{WalMOGWSW07,BatHB13}).  The total mass of Fornax is estimated to be $M_{\rm tot} \simeq 3 \times 10^9 \,M_\odot$, while the mass in stars is only about $M_* \simeq 3 \times 10^7 \, M_\odot$.  The total number of stars in Fornax is approximately $N_* \simeq 5 \times 10^7$, and the dark-matter density is about $\rho_{\rm DM} \simeq 5 \times 10^{-24}\,\mbox{g/cm}^3$ {\color{red} ref?}.  In the following we will adopt these numbers as a concrete example.

%\begin{align}
%M_{\rm tot} & \simeq 3 \times 10^9 M_\odot \\
%M_* & \simeq 3 \times 10^7 M_\odot \\
%N_* & \simeq 5 \times 10^7 \\
%\rho_{\rm DM} & \simeq 5 \times 10^{-24}\,\mbox{g/cm}^3 \\
%v_\infty & \simeq 10\,\mbox{km/s}
%\end{align}

%Number of ``normal" stellar-mass black holes: estimate from Salpeter IMF suggests that about 0.1\% of stars are sufficiently massive to lead to black hole formation, so crudely estimate
%\begin{equation}
%N_{\rm BH}^{\rm SN} \simeq 10^{-3} N_* \simeq 5 \times 10^4.
%\end{equation}

We first estimate the number of PBHs in a typical dwarf galaxy in the {\em direct scenario}.  Specifically, we assume that most of the dwarf galaxy's mass is in dark matter -- consistent with the above numbers -- and that a fraction $f_{\rm PBH}$ of the dark matter mass is in the form of PBHs.   We further assume a ``monochromatic" mass distribution of the PBHs, i.e.~that PBHs have a relatively narrow mass distribution around an average value $M_{\rm PBH}$.  In the direct scenario we take this value to be $M_{\rm PBH} \simeq 10^{-1} M_\odot$, which is in window C and similar to a typical low-mass black hole candidate, in which case we can estimate the number of black holes per dwarf galaxy to be
\begin{equation}
N_{\rm BH,dgal}^{\rm dir} \simeq f_{\rm PBH} \frac{M_{\rm tot}}{M_{\rm PBH}} \simeq 10^9.
\end{equation}
 Here we assumed $f_{\rm PBH} \simeq 0.1$ in the last estimate, which is consistent with the dark matter constraints in window C discussed above.

A Salpeter Initial Mass Function (IMF) suggests that around 0.1\% of all stars have masses greater than 20 $M_\odot$, which we crudely assume to produce a stellar-mass black hole in supernova collapse.  For the Milky Way, with around $10^{11}$ stars, this suggests about $10^8$ such black holes.  Remarkably, then, a single dwarf galaxy could host more PBHs with masses around $0.1 M_\odot$ than the Milky Way Galaxy produces stellar-mass black holes.  Using a similar estimate one expects a typical dwarf galaxy to host around $5 \times 10^4$ stellar-mass black holes. 

We next turn to the {\em indirect scenario}, in which we assume that PBHs are formed with a mass significantly smaller than the observed black-hole mass, $M_{\rm PBH} \ll M_{\rm obs} \simeq 0.1 \, M_{\odot}$.  The density of these PBHs can then be estimated from
\begin{equation} \label{n}
n = f_{\rm PBH} \frac{\rho_{\rm DM}}{M_{\rm PBH}},
\end{equation}
and their rate of collisions with an individual star from
\begin{equation} \label{collision_rate}
\dot {\mathcal N} \simeq \sigma v_\infty n.
\end{equation}
Here $\sigma$ is the cross-section for the collisions, which can be computed from
\begin{equation} \label{sigma}
\sigma = \pi R_*^2 \left(
1 + \left( \frac{v_{\rm esc}}{v_\infty} \right)^2 \right)
\end{equation}
(see, e.g., \cite{AbrBBGJQ09}), where
\begin{equation} \label{v_esc}
v_{\rm esc} = \left( \frac{2 G M_*}{R_*} \right)^{1/2} 
\end{equation}
is the stellar escape speed (the subscript $*$ denotes stellar quantities).  For typical stars the second term on the right-hand side of (\ref{sigma}) dominates, and we will therefore neglect the first term.  The average relative speed (at infinite separation) $v_\infty$ in (\ref{collision_rate}) is related to the (one-dimensional) line-of-sight dispersion velocity $\sigma$ by $v_\infty = \sqrt{2} \, \sigma_{\rm 3D} = \sqrt{6}\, \sigma$, where $\sigma_{\rm 3D} = \sqrt{3} \, \sigma$ is the three-dimensional velocity dispersion.

The initial collision of a PBH with a star will lead to gravitational capture only if a sufficient amount of energy is dissipated during that first transit.  Specifically, the dissipated energy $E_{\rm diss}$, which can be estimated from hydrodynamic friction, has to be greater than the kinetic energy at large separation, $K_\infty \simeq M_{\rm PBH} v_\infty^2 / 2$, for the PBH to remain bound to the star.  This requirement leads to a minimum PBH mass for gravitational capture
\begin{align} \label{M_min}
    M_{\rm PBH}^{\rm min} & \simeq \frac{1}{3 \ln \Lambda} \left( \frac{v_\infty}{v_{\rm esc}} \right)^2 M_* \nonumber \\
    & \simeq 8.7 \times 10^{-6} M_\odot 
    \left( \frac{R_*}{R_\odot} \right)
    \left( \frac{v_\infty}{10 \, \mbox{km/s}} \right)^2
\end{align}
(see Eq.~(14) in \cite{BauS25c}), where we have taken the Coulomb logarithm to be around $\ln \Lambda \simeq 10$ in the second estimate.  Since dwarf stars have radii smaller than the solar radius $R_\odot$, $R_* \simeq 0.1 R_\odot -0.5 R_{\odot}$, we see that PBHs in the mass window B, i.e.~with masses around $10^{-6} M_{\odot}$, could be captured by such stars in environments with relatively low dispersion speeds -- e.g., dwarf galaxies.  As long as $M_{\rm PBH} \ll M_*$ we also assume that at most very little of the stellar matter is ejected in the collisions and/or tidal interactions (see, e.g., \cite{BauS24d,BauS25a} for numerical simulations, albeit for collisions with different types of stars).

Once caught gravitationally, the PBH may re-enter and leave the star many more times, before finally remaining inside the star and settling to its center (see, e.g., \cite{AbrBW18,BauS24c} for illustrations in the context of PBH capture by neutron stars).  The timescale for this swallowing of the PBH by the star depends sensitively on the initial relative speed $v_\infty$ as well as the mass ratio $M_{\rm PBH}/M_*$, because they determine how many transits are required for the PBH to be swallowed, and how much time the PBH spends orbiting outside the star between transits (see \cite{BauS24c} for PBH capture by neutron stars).  We expect, however, that this timescale is much shorter than the lifetime of a dwarf star, which exceeds the age of the Universe $t_{\rm U} = 13.8 \times 10^9 \, \mbox{yrs} = 4.35 \times 10^{17} \, \mbox{s}$.

Once inside the star, the PBH accretes stellar material.  The rate of this accretion, and hence the timescale for the complete consumption of the star, can be estimated from the Bondi accretion rate 
\begin{equation} \label{bondi}
\dot M_{\rm BH} = \frac{4 \pi \lambda G^2 M_{\rm BH}^2 \rho}{a^3}
\end{equation}
(see \cite{Bon52}, as well as \cite{ShaT83} for a textbook treatment) as follows.  In (\ref{bondi}), $M_{\rm BH}$ is the mass of the accreting black hole, $\lambda$ is a dimensionless accretion eigenvalue of order unity (see Table 14.1 in \cite{ShaT83}), $\rho$ is the density of the accreted fluid in the neighborhood of the black hole capture radius $r_c \sim GM_{\rm BH} / a^2 \ll R_*$, and $a$ the sound speed.  Adopting values at the stellar center, for example, we have
\begin{equation} \label{rho_c}
\rho_c \simeq \delta \frac{3 M_*}{4 \pi R_*^3}, 
\end{equation}
where $\delta \equiv \rho_c / \bar \rho$ is the central condensation (and $\bar \rho = 3 M / (4 \pi R^3)$ the average density), and
\begin{equation} \label{a_c}
a_c \simeq \left( \frac{\Gamma G M_*}{R_*} \right)^{1/2},
\end{equation}
where $\Gamma$ is the adiabatic index.  Inserting (\ref{rho_c}) and (\ref{a_c}) into (\ref{bondi}) we obtain a differential equation for the growth of the black hole mass $M_{\rm BH}$.  Integrating this equation from $M_{\rm BH}^{\rm init} = M_{\rm PBH}$ to $M_{\rm BH}^{\rm fin} \simeq M* \gg M_{\rm PBH}$ yields the total accretion timescale,
\begin{equation} \label{tau_acc_1}
\tau_{\rm acc} \simeq \frac{\Gamma^{3/2}}{3 \lambda \delta} \left( \frac{M_*}{M_{\rm PBH}} \right) \left( \frac{R_*^3}{G M_*} \right)^{1/2}
\end{equation}
(see Eq.~(36) in \cite{RicBS21}, where it was also shown that this is an excellent approximation even when the changing structure of the star during consumption is taken into account).
Since the dwarf stars considered here are convective, their structure is well approximated by those of a $\Gamma = 5/3$ polytrope corresponding to a polytropic index $n = 1/(\Gamma - 1) = 3/2$), for which $\delta \simeq 5.99$ (e.g.~Table 4 in \cite{Cha67}) and $\lambda = 1/4$ (see Exercise 14.7 in \cite{ShaT83}).  Further estimating the dynamical timescale 
\begin{equation}
\tau_{\rm dyn} \simeq \left(\frac{R_*^3}{GM_*} \right)^{1/2} \simeq 1.6 \times 10^3\,\mbox{s} \left( \frac{R_*}{R_\odot} \right)^{3/2} \left( \frac{M_*}{M_\odot} \right)^{-1/2}
\end{equation}
we may rewrite (\ref{tau_acc_1}) as
\begin{align}
\tau_{\rm acc} & \simeq 0.48 \left( \frac{M_*}{M_{\rm PBH}} \right) \tau_{\rm dyn} \nonumber \\ 
& \simeq 760\, \mbox{s} \left( \frac{M_*}{M_{\rm PBH}} \right) \left( \frac{R_*}{R_\odot} \right)^{3/2} \left( \frac{M_*}{M_\odot} \right)^{-1/2}.
\end{align}
For $M_{\rm PBH} \ll M_*$ we see that the accretion timescale is much longer than the dynamical timescale, but also shorter than
the stellar lifetime for all PBHs with mass $M_{\rm PBH} \gtrsim 
2 \times 10^{-15} M_{\odot}$. Dynamical simulations of the accretion of a star by an ``endoparasitic" black hole, confirming the above estimates, have been presented in \cite{EasL19,RicBS21,BauS24d,BauS25a}.  In the following we therefore assume that an initial collision of a PBH with mass $M_{\rm PBH}^{\rm min} < M_{\rm PBH} \ll M_*$  will lead to the entire star being accreted by the black hole on timescales shorter than the age of the Universe, and will result in a black hole of mass $M_{\rm BH} \simeq M_*$.

\begin{table*}[t]
    \centering
    \begin{tabular}{c||c|c|c|c||c|c|c|c}
          & \multicolumn{4}{c||}{Simon \cite{Sim19}} &  
         \multicolumn{4}{c}{Walter~{\it et.al.} \cite{WalMOGWSW07}}
         \\
         Galaxy & $v_\infty$ [km/s] & $\rho_{\rm DM}$ [$M_\odot$/pc$^3$] & $\rho_{\rm DM}$  [g/cm$^3$] & ${\mathcal N}_{\rm tot}/f_{\rm PBH}$ & $v_\infty$ [km/s] & $\rho_{\rm DM}$ [$M_\odot$/pc$^3$] & $\rho_{\rm DM}$ [g/cm$^3$] & ${\mathcal N}_{\rm tot}/f_{\rm PBH}$  \\
         \hline
         Carina I & 16.2 & 0.042 & $2.9 \times 10^{-24}$ & $2.5 \times 10^{-4}$ & 17.2 & 0.2 & $1.4\times 10^{-23}$ & $1.1\times 10^{-3}$ \\
         Carina II & 8.3 & 0.13 & $8.8 \times 10^{-24}$ & $1.5 \times 10^{-3}$ & -- & -- & -- & -- \\
         Draco I & 22.3 & 0.15 & $9.9\times 10^{-24}$ & $6.4 \times 10^{-4}$ & 24.5 & 0.4 & $2.7\times 10^{-23}$ & $1.6 \times 10^{-3}$ \\
         Draco II & $<14.4$ & 9.0 & $6.2\times 10^{-22}$ & $>6.1 \times 10^{-2}$ & -- & -- & -- & -- \\
         Fornax & 28.7 & 0.02 & $1.4\times 10^{-24}$ & $7.0\times 10^{-5}$ & 29.4 & 0.3 & $2.1\times 10^{-23}$ & $1.0\times 10^{-3}$ \\
         Leo II & 18.1 & 0.18 & $1.2\times 10^{-23}$ & $9.5 \times 10^{-4}$ & 17.1 & 0.3 & $2.1\times 10^{-23}$ & $1.7 \times 10^{-3}$ \\
         Triangulum II & $<8.3$ & 4.22 & $2.9\times 10^{-22}$ & $>4.9 \times 10^{-2}$ & -- & -- & -- & -- \\
         Segue I & 9.0 & 2.22 & $1.5\times 10^{-24}$ & $2.4 \times 10^{-2}$ & -- & -- & -- & -- \\
         Tucana III & $<2.9$ & 0.10 & $6.7\times 10^{-24}$ & $>3.3 \times 10^{-3}$
   \end{tabular}
    \caption{Sample data for a selection of dwarf galaxies using data from both Simon \cite{Sim19} and -- where available -- Walter {\it et.al.}~\cite{WalMOGWSW07}.  For each galaxy we list the expectation value for the relative speed $v_\infty = \sqrt{2}\, \sigma_{\rm 3D} = \sqrt{6} \,\sigma$ (where $\sigma_{\rm 3D} = \sqrt{3} \,\sigma$ is the three-dimensional velocity dispersion, and $\sigma$ the line-of-sight velocity dispersion), the dark-matter density $\rho_{\rm DM}$, and the total number of collisions ${\mathcal N}_{\rm tot}$ expected between an individual dwarf star (with mass $M_* \simeq 0.1 M_\odot$ and radius $R_* \simeq 0.1 R_\odot$) and a PBH of mass $M_{\rm PBH} \simeq 10^{-6} M_\odot$ , see Eq.~(\ref{Ntot_long}).  For the data of Simon \cite{Sim19}, $v_\infty$ and $\rho_{\rm DM}$ can be computed from the values of $\sigma$
    and the (two-dimensional) half-light-radius $R_{1/2}$ listed in their Table I.  Specifically, we use their Eq.~(1) to compute the half-light mass $M_{1/2}$ as well as the half-light volume $V_{1/2} = 4 \pi r_{1/2}^3/3$ (where $r_{1/2} = 4 R_{1/2}/3$ is the three-dimensional half-light radius), and finally $\rho_{\rm DM} = M_{1/2}/V_{1/2}$, assuming that the mass is dominated by dark matter (see also the data listed in Table I of \cite{EssT23}).  For the data of \cite{WalMOGWSW07} we estimate $\sigma$ from the left panels in their Fig.~2, and estimate $\rho_{\rm DM}$ at a distance of $r = 100\,\mbox{pc}$ from the galaxy centers from the right panel of their Fig.~2. }
    \label{tab:galaxies}
\end{table*}

\begin{widetext}
The (average) total number of collisions ${\mathcal N}_{\rm tot}$ between an individual star and PBHs can now be estimated by multiplying the collision rate (\ref{collision_rate}) with the age of the star.  For massive stars, this age can be estimated from the stellar main-sequence lifetime (see our treatment in \cite{BauS25c}), but for the low-mass dwarf stars considered here this lifetime would be longer than the age of the Universe $t_{\rm U}$ ($\sim$ the age of the Milky Way).  We therefore multiply (\ref{collision_rate}) with $t_{\rm U}$ and insert (\ref{n}), (\ref{sigma}), and (\ref{v_esc}) to obtain
\begin{align} \label{Ntot_long}
{\mathcal N}_{\rm tot} & \simeq
2 \pi f_{\rm PBH} \frac{M_*}{M_{\rm PBH}} \frac{G \rho_{\rm DM} R_* t_{\rm U}}{v_\infty} \nonumber \\
& \simeq 10^{-7} f_{\rm PBH} \left( \frac{M_*}{M_\odot} \right) \left( \frac{R_*}{R_\odot} \right) \left( \frac{v_\infty}{10\,\mbox{km/s}} \right)^{-1} \left( \frac{M_{\rm PBH}}{M_\odot} \right)^{-1}  \left( \frac{\rho_{\rm DM}}{7 \times 10^{-24} \, \mbox{g/cm}^3} \right).
\end{align}
%\begin{equation}
%{\mathcal N}_{\rm tot} \simeq 10^{-8} \left( \frac{M_*}{M_\odot} \right)^{1-\alpha}  \left( \frac{R_*}{R_\odot} \right) \left( \frac{v_\infty}{10^6\,\mbox{cm/s}} \right)^{-1} \left( \frac{M_{\rm PBH}}{M_\odot} \right)^{-1} f_{\rm PBH} \frac{\rho_{\rm DM}}{\rho_{\rm DM}^{\rm loc}}
%\end{equation}
\end{widetext}
Adopting $R_* \simeq 0.1 R_\odot$ for a dwarf star of mass $M_* \simeq 0.1 M_\odot$, and using the above fiducial values for $v_\infty$ and $\rho_{\rm DM}$, we have 
\begin{equation} \label{Ntot}
{\mathcal N}_{\rm tot} \simeq 10^{-3} f_{\rm PBH}
\end{equation}
for a typical dwarf star and for $M_{\rm PBH} \simeq 10^{-6} M_\odot$.\footnote{Alternatively, Eq.~(\ref{Ntot}) can be derived by considering the probability of induced collapse for a dwarf star, $P=t_{\rm U}/t_{\rm coll} = (\sigma v_\infty n) t_{\rm U} = {\mathcal N}_{\rm tot}$, where $t_{\rm coll}$ is the collision time.}  In Table \ref{tab:galaxies} we list specific values of ${\mathcal N}_{\rm tot}$ for a sample of dwarf galaxies.  There are significant differences between different galaxies, and uncertainties in the dark-matter density $\rho_{\rm DM}$ in particular also translate into uncertainties in ${\mathcal N}_{\rm tot}$.  Crudely speaking, however, the values listed in Table \ref{tab:galaxies} average around a value similar to (\ref{Ntot}), which we will therefore adopt in the following.
According to the Salpeter IMF, about 70\% of stars are dwarf stars, so the total number of black holes formed in this indirect ``induced collapse" scenario is
\begin{equation} \label{NBHDG}
N_{\rm BH,dgal}^{\rm ind} \simeq 0.7 N_* {\mathcal N}_{\rm tot}  \simeq 3 \times 10^4 f_{\rm PBH} \simeq 3 \times 10^3
\end{equation}
per dwarf galaxy, assuming $N_* \simeq 5 \times 10^7$ (note, however, that some of the ultra-faint galaxies listed in Table \ref{tab:galaxies} contain significantly fewer stars).  We also assumed $f_{\rm PBH} = 0.1$ again in the last step, consistent with the constraints in window B discussed above. Further assuming that there are around 50 dwarf galaxies per Milky Way-type galaxy, the number of subsolar mass black holes produced in the indirect PBH scenario in the dwarf galaxies alone could reach
\begin{equation}
N^{\rm ind}_{\rm BH,gal} \simeq 10^5
\end{equation}
per Milky Way-type galaxy.  While the higher dispersion speed in the Galaxy itself makes the capture process less likely there, outliers in the distribution of relative speeds $v_\infty$ may contribute more subsolar mass black holes.  

While a future confirmed detection of a subsolar mass black hole with mass $M_{\rm obs} < M_\odot$ would suggest a primordial origin, we therefore conclude that it would not necessarily imply that PBHs are formed with masses around $M_{\rm obs}$ -- as is assumed in what we refer to as the ``direct PBH scenario".  In an alternative ``indirect PBH scenario", smaller PBHs could collide with dwarf stars and induce collapse to form subsolar mass black holes around $0.1\,M_\odot$.  Our crude estimates suggest that the number of subsolar mass black holes that can be formed in this way -- around $10^5$ or so per Milky Way-type galaxy -- is dramatically smaller than the number allowed in the direct PBH scenario, and also only about 0.1\% of the $10^8$ or so stellar-mass black holes formed in supernovae.  However, this number may still be large enough to explain rare events involving subsolar mass black holes, especially if other constraints rule out large populations of PBHs with masses around $M_{\rm obs}$, and hence the direct PBH scenario (see, e.g., \cite{YuaH24} for a discussion).  

While it is unlikely that a gravitational-wave observation alone could distinguish sub-solar mass black holes formed in the direct scenario from those formed in the indirect scenario, it may be possible to discern the two scenarios using other observations.  In particular, the authors of \cite{EssDT24} note that PBH capture rates depend on stellar mass, so that the indirect scenario would leave a potentially observable imprint in the mass distribution, especially in ultra-faint dwarf galaxies.

Finally we note that we consider PBH capture by isolated stars only, and ignore three-body interactions that may complicate the capture process for stars in close binaries.  Allowing for such effects would indicate that our estimates above serve as upper limits on the indirect scenario for forming subsolar mass black holes. 

\acknowledgments

We would like to thank Gabriela Gonz\'alez, Nicolas Esser, and Peter Tinyakov for helpful comments.  This work was supported in parts by National Science Foundation (NSF) grant PHY-2308821 to Bowdoin College and NSF grant PHY-2308242 to the University of Illinois at Urbana-Champaign. 

% \bibliography{references}

\begin{thebibliography}{39}%
\makeatletter
\providecommand \@ifxundefined [1]{%
 \@ifx{#1\undefined}
}%
\providecommand \@ifnum [1]{%
 \ifnum #1\expandafter \@firstoftwo
 \else \expandafter \@secondoftwo
 \fi
}%
\providecommand \@ifx [1]{%
 \ifx #1\expandafter \@firstoftwo
 \else \expandafter \@secondoftwo
 \fi
}%
\providecommand \natexlab [1]{#1}%
\providecommand \enquote  [1]{``#1''}%
\providecommand \bibnamefont  [1]{#1}%
\providecommand \bibfnamefont [1]{#1}%
\providecommand \citenamefont [1]{#1}%
\providecommand \href@noop [0]{\@secondoftwo}%
\providecommand \href [0]{\begingroup \@sanitize@url \@href}%
\providecommand \@href[1]{\@@startlink{#1}\@@href}%
\providecommand \@@href[1]{\endgroup#1\@@endlink}%
\providecommand \@sanitize@url [0]{\catcode `\\12\catcode `\$12\catcode
  `\&12\catcode `\#12\catcode `\^12\catcode `\_12\catcode `\%12\relax}%
\providecommand \@@startlink[1]{}%
\providecommand \@@endlink[0]{}%
\providecommand \url  [0]{\begingroup\@sanitize@url \@url }%
\providecommand \@url [1]{\endgroup\@href {#1}{\urlprefix }}%
\providecommand \urlprefix  [0]{URL }%
\providecommand \Eprint [0]{\href }%
\providecommand \doibase [0]{https://doi.org/}%
\providecommand \selectlanguage [0]{\@gobble}%
\providecommand \bibinfo  [0]{\@secondoftwo}%
\providecommand \bibfield  [0]{\@secondoftwo}%
\providecommand \translation [1]{[#1]}%
\providecommand \BibitemOpen [0]{}%
\providecommand \bibitemStop [0]{}%
\providecommand \bibitemNoStop [0]{.\EOS\space}%
\providecommand \EOS [0]{\spacefactor3000\relax}%
\providecommand \BibitemShut  [1]{\csname bibitem#1\endcsname}%
\let\auto@bib@innerbib\@empty
%</preamble>
\bibitem [{\citenamefont {{The LIGO Scientific Collaboration}}\ and\
  \citenamefont {{the Virgo Collaboration}}(2016)}]{LIGO_GW150914}%
  \BibitemOpen
  \bibfield  {author} {\bibinfo {author} {\bibnamefont {{The LIGO Scientific
  Collaboration}}}\ and\ \bibinfo {author} {\bibnamefont {{the Virgo
  Collaboration}}},\ }\bibfield  {title} {\bibinfo {title} {{Observation of
  Gravitational Waves from a Binary Black Hole Merger}},\ }\href
  {https://doi.org/10.1103/PhysRevLett.116.061102} {\bibfield  {journal}
  {\bibinfo  {journal} {\prl}\ }\textbf {\bibinfo {volume} {116}},\ \bibinfo
  {eid} {061102} (\bibinfo {year} {2016})},\ \Eprint
  {https://arxiv.org/abs/1602.03837} {arXiv:1602.03837 [gr-qc]} \BibitemShut
  {NoStop}%
\bibitem [{\citenamefont {{Suwa}}\ \emph {et~al.}(2018)\citenamefont {{Suwa}},
  \citenamefont {{Yoshida}}, \citenamefont {{Shibata}}, \citenamefont
  {{Umeda}},\ and\ \citenamefont {{Takahashi}}}]{SuwYSUT18}%
  \BibitemOpen
  \bibfield  {author} {\bibinfo {author} {\bibfnamefont {Y.}~\bibnamefont
  {{Suwa}}}, \bibinfo {author} {\bibfnamefont {T.}~\bibnamefont {{Yoshida}}},
  \bibinfo {author} {\bibfnamefont {M.}~\bibnamefont {{Shibata}}}, \bibinfo
  {author} {\bibfnamefont {H.}~\bibnamefont {{Umeda}}},\ and\ \bibinfo {author}
  {\bibfnamefont {K.}~\bibnamefont {{Takahashi}}},\ }\bibfield  {title}
  {\bibinfo {title} {{On the minimum mass of neutron stars}},\ }\href
  {https://doi.org/10.1093/mnras/sty2460} {\bibfield  {journal} {\bibinfo
  {journal} {\mnras}\ }\textbf {\bibinfo {volume} {481}},\ \bibinfo {pages}
  {3305} (\bibinfo {year} {2018})},\ \Eprint {https://arxiv.org/abs/1808.02328}
  {arXiv:1808.02328 [astro-ph.HE]} \BibitemShut {NoStop}%
\bibitem [{\citenamefont {{Doroshenko}}\ \emph {et~al.}(2022)\citenamefont
  {{Doroshenko}}, \citenamefont {{Suleimanov}}, \citenamefont
  {{P{\"u}hlhofer}},\ and\ \citenamefont {{Santangelo}}}]{DorSPS22}%
  \BibitemOpen
  \bibfield  {author} {\bibinfo {author} {\bibfnamefont {V.}~\bibnamefont
  {{Doroshenko}}}, \bibinfo {author} {\bibfnamefont {V.}~\bibnamefont
  {{Suleimanov}}}, \bibinfo {author} {\bibfnamefont {G.}~\bibnamefont
  {{P{\"u}hlhofer}}},\ and\ \bibinfo {author} {\bibfnamefont {A.}~\bibnamefont
  {{Santangelo}}},\ }\bibfield  {title} {\bibinfo {title} {{A strangely light
  neutron star within a supernova remnant}},\ }\href
  {https://doi.org/10.1038/s41550-022-01800-1} {\bibfield  {journal} {\bibinfo
  {journal} {Nature Astronomy}\ }\textbf {\bibinfo {volume} {6}},\ \bibinfo
  {pages} {1444} (\bibinfo {year} {2022})}\BibitemShut {NoStop}%
\bibitem [{\citenamefont {{Crescimbeni}}(2025)}]{Cre25}%
  \BibitemOpen
  \bibfield  {author} {\bibinfo {author} {\bibfnamefont {F.}~\bibnamefont
  {{Crescimbeni}}},\ }\bibfield  {title} {\bibinfo {title} {{Can we identify
  primordial black holes? The role of subsolar gravitational wave events}},\
  }\href {https://doi.org/10.48550/arXiv.2511.01051} {\bibfield  {journal}
  {\bibinfo  {journal} {arXiv e-prints}\ ,\ \bibinfo {eid} {arXiv:2511.01051}}
  (\bibinfo {year} {2025})},\ \Eprint {https://arxiv.org/abs/2511.01051}
  {arXiv:2511.01051 [gr-qc]} \BibitemShut {NoStop}%
\bibitem [{\citenamefont {{Aggarwal}}\ \emph {et~al.}(2025)\citenamefont
  {{Aggarwal}}, \citenamefont {{Aguiar}}, \citenamefont {{Blas}}, \citenamefont
  {{Bauswein}}, \citenamefont {{Cella}}, \citenamefont {{Clesse}},
  \citenamefont {{Cruise}}, \citenamefont {{Domcke}}, \citenamefont {{Ellis}},
  \citenamefont {{Figueroa}}, \citenamefont {{Franciolini}}, \citenamefont
  {{Garcia-Cely}}, \citenamefont {{Geraci}}, \citenamefont {{Goryachev}},
  \citenamefont {{Grote}}, \citenamefont {{Hindmarsh}}, \citenamefont {{Ito}},
  \citenamefont {{Kopp}}, \citenamefont {{Mook Lee}}, \citenamefont
  {{Martineau}}, \citenamefont {{McDonald}}, \citenamefont {{Muia}},
  \citenamefont {{Mukund}}, \citenamefont {{Ottaway}}, \citenamefont
  {{Peloso}}, \citenamefont {{Peters}}, \citenamefont {{Quevedo}},
  \citenamefont {{Ricciardone}}, \citenamefont {{Ringwald}}, \citenamefont
  {{Steinlechner}}, \citenamefont {{Steinlechner}}, \citenamefont {{Sun}},
  \citenamefont {{Tamarit}}, \citenamefont {{Tobar}}, \citenamefont
  {{Torrenti}}, \citenamefont {{{\"U}nal}},\ and\ \citenamefont
  {{White}}}]{Aggetal25}%
  \BibitemOpen
  \bibfield  {author} {\bibinfo {author} {\bibfnamefont {N.}~\bibnamefont
  {{Aggarwal}}}, \bibinfo {author} {\bibfnamefont {O.~D.}\ \bibnamefont
  {{Aguiar}}}, \bibinfo {author} {\bibfnamefont {D.}~\bibnamefont {{Blas}}},
  \bibinfo {author} {\bibfnamefont {A.}~\bibnamefont {{Bauswein}}}, \bibinfo
  {author} {\bibfnamefont {G.}~\bibnamefont {{Cella}}}, \bibinfo {author}
  {\bibfnamefont {S.}~\bibnamefont {{Clesse}}}, \bibinfo {author}
  {\bibfnamefont {A.~M.}\ \bibnamefont {{Cruise}}}, \bibinfo {author}
  {\bibfnamefont {V.}~\bibnamefont {{Domcke}}}, \bibinfo {author}
  {\bibfnamefont {S.}~\bibnamefont {{Ellis}}}, \bibinfo {author} {\bibfnamefont
  {D.~G.}\ \bibnamefont {{Figueroa}}}, \bibinfo {author} {\bibfnamefont
  {G.}~\bibnamefont {{Franciolini}}}, \bibinfo {author} {\bibfnamefont
  {C.}~\bibnamefont {{Garcia-Cely}}}, \bibinfo {author} {\bibfnamefont
  {A.}~\bibnamefont {{Geraci}}}, \bibinfo {author} {\bibfnamefont
  {M.}~\bibnamefont {{Goryachev}}}, \bibinfo {author} {\bibfnamefont
  {H.}~\bibnamefont {{Grote}}}, \bibinfo {author} {\bibfnamefont
  {M.}~\bibnamefont {{Hindmarsh}}}, \bibinfo {author} {\bibfnamefont
  {A.}~\bibnamefont {{Ito}}}, \bibinfo {author} {\bibfnamefont
  {J.}~\bibnamefont {{Kopp}}}, \bibinfo {author} {\bibfnamefont
  {S.}~\bibnamefont {{Mook Lee}}}, \bibinfo {author} {\bibfnamefont
  {K.}~\bibnamefont {{Martineau}}}, \bibinfo {author} {\bibfnamefont
  {J.}~\bibnamefont {{McDonald}}}, \bibinfo {author} {\bibfnamefont
  {F.}~\bibnamefont {{Muia}}}, \bibinfo {author} {\bibfnamefont
  {N.}~\bibnamefont {{Mukund}}}, \bibinfo {author} {\bibfnamefont
  {D.}~\bibnamefont {{Ottaway}}}, \bibinfo {author} {\bibfnamefont
  {M.}~\bibnamefont {{Peloso}}}, \bibinfo {author} {\bibfnamefont
  {K.}~\bibnamefont {{Peters}}}, \bibinfo {author} {\bibfnamefont
  {F.}~\bibnamefont {{Quevedo}}}, \bibinfo {author} {\bibfnamefont
  {A.}~\bibnamefont {{Ricciardone}}}, \bibinfo {author} {\bibfnamefont
  {A.}~\bibnamefont {{Ringwald}}}, \bibinfo {author} {\bibfnamefont
  {J.}~\bibnamefont {{Steinlechner}}}, \bibinfo {author} {\bibfnamefont
  {S.}~\bibnamefont {{Steinlechner}}}, \bibinfo {author} {\bibfnamefont
  {S.}~\bibnamefont {{Sun}}}, \bibinfo {author} {\bibfnamefont
  {C.}~\bibnamefont {{Tamarit}}}, \bibinfo {author} {\bibfnamefont {M.~E.}\
  \bibnamefont {{Tobar}}}, \bibinfo {author} {\bibfnamefont {F.}~\bibnamefont
  {{Torrenti}}}, \bibinfo {author} {\bibfnamefont {C.}~\bibnamefont
  {{{\"U}nal}}},\ and\ \bibinfo {author} {\bibfnamefont {G.}~\bibnamefont
  {{White}}},\ }\bibfield  {title} {\bibinfo {title} {{Challenges and
  Opportunities of Gravitational Wave Searches above 10 kHz}},\ }\href
  {https://doi.org/10.48550/arXiv.2501.11723} {\bibfield  {journal} {\bibinfo
  {journal} {arXiv e-prints}\ ,\ \bibinfo {eid} {arXiv:2501.11723}} (\bibinfo
  {year} {2025})},\ \Eprint {https://arxiv.org/abs/2501.11723}
  {arXiv:2501.11723 [gr-qc]} \BibitemShut {NoStop}%
\bibitem [{\citenamefont {{Miller}}\ and\ \citenamefont
  {{Pierini}}(2025)}]{MilP25}%
  \BibitemOpen
  \bibfield  {author} {\bibinfo {author} {\bibfnamefont {A.~L.}\ \bibnamefont
  {{Miller}}}\ and\ \bibinfo {author} {\bibfnamefont {L.}~\bibnamefont
  {{Pierini}}},\ }\bibfield  {title} {\bibinfo {title} {{BinaryGFH-v2: Improved
  method to search for gravitational waves from sub-solar-mass, ultra-compact
  binaries using the Generalized Frequency-Hough Transform}},\ }\href@noop {}
  {\bibfield  {journal} {\bibinfo  {journal} {arXiv e-prints}\ ,\ \bibinfo
  {eid} {arXiv:2512.10539}} (\bibinfo {year} {2025})},\ \Eprint
  {https://arxiv.org/abs/2512.10539} {arXiv:2512.10539 [gr-qc]} \BibitemShut
  {NoStop}%
\bibitem [{\citenamefont {{Carr}}\ \emph {et~al.}(2024)\citenamefont {{Carr}},
  \citenamefont {{Clesse}}, \citenamefont {{Garc{\'\i}a-Bellido}},
  \citenamefont {{Hawkins}},\ and\ \citenamefont {{K{\"u}hnel}}}]{CarCGHK24}%
  \BibitemOpen
  \bibfield  {author} {\bibinfo {author} {\bibfnamefont {B.~J.}\ \bibnamefont
  {{Carr}}}, \bibinfo {author} {\bibfnamefont {S.}~\bibnamefont {{Clesse}}},
  \bibinfo {author} {\bibfnamefont {J.}~\bibnamefont {{Garc{\'\i}a-Bellido}}},
  \bibinfo {author} {\bibfnamefont {M.~R.~S.}\ \bibnamefont {{Hawkins}}},\ and\
  \bibinfo {author} {\bibfnamefont {F.}~\bibnamefont {{K{\"u}hnel}}},\
  }\bibfield  {title} {\bibinfo {title} {{Observational evidence for primordial
  black holes: A positivist perspective}},\ }\href
  {https://doi.org/10.1016/j.physrep.2023.11.005} {\bibfield  {journal}
  {\bibinfo  {journal} {\physrep}\ }\textbf {\bibinfo {volume} {1054}},\
  \bibinfo {pages} {1} (\bibinfo {year} {2024})},\ \Eprint
  {https://arxiv.org/abs/2306.03903} {arXiv:2306.03903 [astro-ph.CO]}
  \BibitemShut {NoStop}%
\bibitem [{\citenamefont {{Phukon}}\ \emph {et~al.}(2021)\citenamefont
  {{Phukon}}, \citenamefont {{Baltus}}, \citenamefont {{Caudill}},
  \citenamefont {{Clesse}}, \citenamefont {{Depasse}}, \citenamefont {{Fays}},
  \citenamefont {{Fong}}, \citenamefont {{Kapadia}}, \citenamefont {{Magee}},\
  and\ \citenamefont {{Tanasijczuk}}}]{Phuetal21}%
  \BibitemOpen
  \bibfield  {author} {\bibinfo {author} {\bibfnamefont {K.~S.}\ \bibnamefont
  {{Phukon}}}, \bibinfo {author} {\bibfnamefont {G.}~\bibnamefont {{Baltus}}},
  \bibinfo {author} {\bibfnamefont {S.}~\bibnamefont {{Caudill}}}, \bibinfo
  {author} {\bibfnamefont {S.}~\bibnamefont {{Clesse}}}, \bibinfo {author}
  {\bibfnamefont {A.}~\bibnamefont {{Depasse}}}, \bibinfo {author}
  {\bibfnamefont {M.}~\bibnamefont {{Fays}}}, \bibinfo {author} {\bibfnamefont
  {H.}~\bibnamefont {{Fong}}}, \bibinfo {author} {\bibfnamefont {S.~J.}\
  \bibnamefont {{Kapadia}}}, \bibinfo {author} {\bibfnamefont {R.}~\bibnamefont
  {{Magee}}},\ and\ \bibinfo {author} {\bibfnamefont {A.~J.}\ \bibnamefont
  {{Tanasijczuk}}},\ }\bibfield  {title} {\bibinfo {title} {{The hunt for
  sub-solar primordial black holes in low mass ratio binaries is open}},\
  }\href {https://doi.org/10.48550/arXiv.2105.11449} {\bibfield  {journal}
  {\bibinfo  {journal} {arXiv e-prints}\ ,\ \bibinfo {eid} {arXiv:2105.11449}}
  (\bibinfo {year} {2021})},\ \Eprint {https://arxiv.org/abs/2105.11449}
  {arXiv:2105.11449 [astro-ph.CO]} \BibitemShut {NoStop}%
\bibitem [{\citenamefont {{Morr{\'a}s}}\ \emph {et~al.}(2023)\citenamefont
  {{Morr{\'a}s}}, \citenamefont {{Nu{\~n}o Siles}}, \citenamefont
  {{Garc{\'\i}a-Bellido}}, \citenamefont {{Ruiz Morales}}, \citenamefont
  {{Men{\'e}ndez-V{\'a}zquez}}, \citenamefont {{Karathanasis}}, \citenamefont
  {{Martinovic}}, \citenamefont {{Phukon}}, \citenamefont {{Clesse}},
  \citenamefont {{Mart{\'\i}nez}},\ and\ \citenamefont
  {{Sakellariadou}}}]{Moretal23}%
  \BibitemOpen
  \bibfield  {author} {\bibinfo {author} {\bibfnamefont {G.}~\bibnamefont
  {{Morr{\'a}s}}}, \bibinfo {author} {\bibfnamefont {J.~F.}\ \bibnamefont
  {{Nu{\~n}o Siles}}}, \bibinfo {author} {\bibfnamefont {J.}~\bibnamefont
  {{Garc{\'\i}a-Bellido}}}, \bibinfo {author} {\bibfnamefont {E.}~\bibnamefont
  {{Ruiz Morales}}}, \bibinfo {author} {\bibfnamefont {A.}~\bibnamefont
  {{Men{\'e}ndez-V{\'a}zquez}}}, \bibinfo {author} {\bibfnamefont
  {C.}~\bibnamefont {{Karathanasis}}}, \bibinfo {author} {\bibfnamefont
  {K.}~\bibnamefont {{Martinovic}}}, \bibinfo {author} {\bibfnamefont {K.~S.}\
  \bibnamefont {{Phukon}}}, \bibinfo {author} {\bibfnamefont {S.}~\bibnamefont
  {{Clesse}}}, \bibinfo {author} {\bibfnamefont {M.}~\bibnamefont
  {{Mart{\'\i}nez}}},\ and\ \bibinfo {author} {\bibfnamefont {M.}~\bibnamefont
  {{Sakellariadou}}},\ }\bibfield  {title} {\bibinfo {title} {{Analysis of a
  subsolar-mass compact binary candidate from the second observing run of
  Advanced LIGO}},\ }\href {https://doi.org/10.1016/j.dark.2023.101285}
  {\bibfield  {journal} {\bibinfo  {journal} {Physics of the Dark Universe}\
  }\textbf {\bibinfo {volume} {42}},\ \bibinfo {eid} {101285} (\bibinfo {year}
  {2023})},\ \Eprint {https://arxiv.org/abs/2301.11619} {arXiv:2301.11619
  [gr-qc]} \BibitemShut {NoStop}%
\bibitem [{\citenamefont {{Prunier}}\ \emph {et~al.}(2024)\citenamefont
  {{Prunier}}, \citenamefont {{Morr{\'a}s}}, \citenamefont {{Siles}},
  \citenamefont {{Clesse}}, \citenamefont {{Garc{\'\i}a-Bellido}},\ and\
  \citenamefont {{Morales}}}]{Pruetal24}%
  \BibitemOpen
  \bibfield  {author} {\bibinfo {author} {\bibfnamefont {M.}~\bibnamefont
  {{Prunier}}}, \bibinfo {author} {\bibfnamefont {G.}~\bibnamefont
  {{Morr{\'a}s}}}, \bibinfo {author} {\bibfnamefont {J.~F.~N.}\ \bibnamefont
  {{Siles}}}, \bibinfo {author} {\bibfnamefont {S.}~\bibnamefont {{Clesse}}},
  \bibinfo {author} {\bibfnamefont {J.}~\bibnamefont {{Garc{\'\i}a-Bellido}}},\
  and\ \bibinfo {author} {\bibfnamefont {E.~R.}\ \bibnamefont {{Morales}}},\
  }\bibfield  {title} {\bibinfo {title} {{Analysis of the subsolar-mass black
  hole candidate SSM200308 from the second part of the third observing run of
  Advanced LIGO-Virgo}},\ }\href {https://doi.org/10.1016/j.dark.2024.101582}
  {\bibfield  {journal} {\bibinfo  {journal} {Physics of the Dark Universe}\
  }\textbf {\bibinfo {volume} {46}},\ \bibinfo {eid} {101582} (\bibinfo {year}
  {2024})},\ \Eprint {https://arxiv.org/abs/2311.16085} {arXiv:2311.16085
  [gr-qc]} \BibitemShut {NoStop}%
\bibitem [{\citenamefont {{Ligo Scientific Collaboration}}\ \emph
  {et~al.}(2025)\citenamefont {{Ligo Scientific Collaboration}}, \citenamefont
  {{VIRGO Collaboration}},\ and\ \citenamefont {{Kagra
  Collaboration}}}]{2025GCN.42650....1L}%
  \BibitemOpen
  \bibfield  {author} {\bibinfo {author} {\bibnamefont {{Ligo Scientific
  Collaboration}}}, \bibinfo {author} {\bibnamefont {{VIRGO Collaboration}}},\
  and\ \bibinfo {author} {\bibnamefont {{Kagra Collaboration}}},\ }\bibfield
  {title} {\bibinfo {title} {{LIGO/Virgo/KAGRA S251112cm: Identification of a
  GW compact binary merger candidate}},\ }\href@noop {} {\bibfield  {journal}
  {\bibinfo  {journal} {GRB Coordinates Network}\ }\textbf {\bibinfo {volume}
  {42650}},\ \bibinfo {pages} {1} (\bibinfo {year} {2025})}\BibitemShut
  {NoStop}%
\bibitem [{\citenamefont {{LVK Collaboration}}()}]{GraceDB}%
  \BibitemOpen
  \bibfield  {author} {\bibinfo {author} {\bibnamefont {{LVK Collaboration}}},\
  }\bibfield  {title} {\bibinfo {title} {{GraceDB Alert S251112cm}},\ }\bibinfo
  {note} {https://gracedb.ligo.org/superevents/S251112cm/}\BibitemShut
  {NoStop}%
\bibitem [{\citenamefont {{Nitz}}\ and\ \citenamefont {{Wang}}(2021)}]{NitW21}%
  \BibitemOpen
  \bibfield  {author} {\bibinfo {author} {\bibfnamefont {A.~H.}\ \bibnamefont
  {{Nitz}}}\ and\ \bibinfo {author} {\bibfnamefont {Y.-F.}\ \bibnamefont
  {{Wang}}},\ }\bibfield  {title} {\bibinfo {title} {{Search for Gravitational
  Waves from the Coalescence of Subsolar-Mass Binaries in the First Half of
  Advanced LIGO and Virgo's Third Observing Run}},\ }\href
  {https://doi.org/10.1103/PhysRevLett.127.151101} {\bibfield  {journal}
  {\bibinfo  {journal} {\prl}\ }\textbf {\bibinfo {volume} {127}},\ \bibinfo
  {eid} {151101} (\bibinfo {year} {2021})},\ \Eprint
  {https://arxiv.org/abs/2106.08979} {arXiv:2106.08979 [astro-ph.HE]}
  \BibitemShut {NoStop}%
\bibitem [{\citenamefont {{LVK Collaboration}}(2023)}]{LVK23}%
  \BibitemOpen
  \bibfield  {author} {\bibinfo {author} {\bibnamefont {{LVK Collaboration}}},\
  }\bibfield  {title} {\bibinfo {title} {{Search for subsolar-mass black hole
  binaries in the second part of Advanced LIGO's and Advanced Virgo's third
  observing run}},\ }\href {https://doi.org/10.1093/mnras/stad588} {\bibfield
  {journal} {\bibinfo  {journal} {\mnras}\ }\textbf {\bibinfo {volume} {524}},\
  \bibinfo {pages} {5984} (\bibinfo {year} {2023})}\BibitemShut {NoStop}%
\bibitem [{\citenamefont {{The LIGO Scientific Collaboration}}\ and\
  \citenamefont {{the Virgo Collaboration}}(2019)}]{LIGO19}%
  \BibitemOpen
  \bibfield  {author} {\bibinfo {author} {\bibnamefont {{The LIGO Scientific
  Collaboration}}}\ and\ \bibinfo {author} {\bibnamefont {{the Virgo
  Collaboration}}},\ }\bibfield  {title} {\bibinfo {title} {{Search for
  Subsolar Mass Ultracompact Binaries in Advanced LIGO's Second Observing
  Run}},\ }\href {https://doi.org/10.1103/PhysRevLett.123.161102} {\bibfield
  {journal} {\bibinfo  {journal} {\prl}\ }\textbf {\bibinfo {volume} {123}},\
  \bibinfo {eid} {161102} (\bibinfo {year} {2019})}\BibitemShut {NoStop}%
\bibitem [{\citenamefont {{Nitz}}\ and\ \citenamefont {{Wang}}(2022)}]{NitW22}%
  \BibitemOpen
  \bibfield  {author} {\bibinfo {author} {\bibfnamefont {A.~H.}\ \bibnamefont
  {{Nitz}}}\ and\ \bibinfo {author} {\bibfnamefont {Y.-F.}\ \bibnamefont
  {{Wang}}},\ }\bibfield  {title} {\bibinfo {title} {{Broad search for
  gravitational waves from subsolar-mass binaries through LIGO and Virgo's
  third observing run}},\ }\href {https://doi.org/10.1103/PhysRevD.106.023024}
  {\bibfield  {journal} {\bibinfo  {journal} {\prd}\ }\textbf {\bibinfo
  {volume} {106}},\ \bibinfo {eid} {023024} (\bibinfo {year} {2022})},\ \Eprint
  {https://arxiv.org/abs/2202.11024} {arXiv:2202.11024 [astro-ph.HE]}
  \BibitemShut {NoStop}%
\bibitem [{\citenamefont {{Yuan}}\ and\ \citenamefont
  {{Huang}}(2024)}]{YuaH24}%
  \BibitemOpen
  \bibfield  {author} {\bibinfo {author} {\bibfnamefont {C.}~\bibnamefont
  {{Yuan}}}\ and\ \bibinfo {author} {\bibfnamefont {Q.-G.}\ \bibnamefont
  {{Huang}}},\ }\bibfield  {title} {\bibinfo {title} {{Primordial black hole
  interpretation in subsolar mass gravitational wave candidate SSM200308}},\
  }\href {https://doi.org/10.1088/1475-7516/2024/09/051} {\bibfield  {journal}
  {\bibinfo  {journal} {\jcap}\ }\textbf {\bibinfo {volume} {2024}},\ \bibinfo
  {eid} {051} (\bibinfo {year} {2024})},\ \Eprint
  {https://arxiv.org/abs/2404.03328} {arXiv:2404.03328 [astro-ph.CO]}
  \BibitemShut {NoStop}%
\bibitem [{\citenamefont {{Carr}}\ and\ \citenamefont
  {{K{\"u}hnel}}(2020)}]{CarK20}%
  \BibitemOpen
  \bibfield  {author} {\bibinfo {author} {\bibfnamefont {B.}~\bibnamefont
  {{Carr}}}\ and\ \bibinfo {author} {\bibfnamefont {F.}~\bibnamefont
  {{K{\"u}hnel}}},\ }\bibfield  {title} {\bibinfo {title} {{Primordial Black
  Holes as Dark Matter: Recent Developments}},\ }\href
  {https://doi.org/10.1146/annurev-nucl-050520-125911} {\bibfield  {journal}
  {\bibinfo  {journal} {Annual Review of Nuclear and Particle Science}\
  }\textbf {\bibinfo {volume} {70}},\ \bibinfo {pages} {355} (\bibinfo {year}
  {2020})},\ \Eprint {https://arxiv.org/abs/2006.02838} {arXiv:2006.02838
  [astro-ph.CO]} \BibitemShut {NoStop}%
\bibitem [{\citenamefont {{Carr}}\ \emph {et~al.}(2021)\citenamefont {{Carr}},
  \citenamefont {{Kohri}}, \citenamefont {{Sendouda}},\ and\ \citenamefont
  {{Yokoyama}}}]{CarKSY21}%
  \BibitemOpen
  \bibfield  {author} {\bibinfo {author} {\bibfnamefont {B.}~\bibnamefont
  {{Carr}}}, \bibinfo {author} {\bibfnamefont {K.}~\bibnamefont {{Kohri}}},
  \bibinfo {author} {\bibfnamefont {Y.}~\bibnamefont {{Sendouda}}},\ and\
  \bibinfo {author} {\bibfnamefont {J.}~\bibnamefont {{Yokoyama}}},\ }\bibfield
   {title} {\bibinfo {title} {{Constraints on primordial black holes}},\ }\href
  {https://doi.org/10.1088/1361-6633/ac1e31} {\bibfield  {journal} {\bibinfo
  {journal} {Reports on Progress in Physics}\ }\textbf {\bibinfo {volume}
  {84}},\ \bibinfo {eid} {116902} (\bibinfo {year} {2021})},\ \Eprint
  {https://arxiv.org/abs/2002.12778} {arXiv:2002.12778 [astro-ph.CO]}
  \BibitemShut {NoStop}%
\bibitem [{\citenamefont {{Abramowicz}}\ \emph {et~al.}(2009)\citenamefont
  {{Abramowicz}}, \citenamefont {{Becker}}, \citenamefont {{Biermann}},
  \citenamefont {{Garzilli}}, \citenamefont {{Johansson}},\ and\ \citenamefont
  {{Qian}}}]{AbrBBGJQ09}%
  \BibitemOpen
  \bibfield  {author} {\bibinfo {author} {\bibfnamefont {M.~A.}\ \bibnamefont
  {{Abramowicz}}}, \bibinfo {author} {\bibfnamefont {J.~K.}\ \bibnamefont
  {{Becker}}}, \bibinfo {author} {\bibfnamefont {P.~L.}\ \bibnamefont
  {{Biermann}}}, \bibinfo {author} {\bibfnamefont {A.}~\bibnamefont
  {{Garzilli}}}, \bibinfo {author} {\bibfnamefont {F.}~\bibnamefont
  {{Johansson}}},\ and\ \bibinfo {author} {\bibfnamefont {L.}~\bibnamefont
  {{Qian}}},\ }\bibfield  {title} {\bibinfo {title} {{No Observational
  Constraints from Hypothetical Collisions of Hypothetical Dark Halo Primordial
  Black Holes with Galactic Objects}},\ }\href
  {https://doi.org/10.1088/0004-637X/705/1/659} {\bibfield  {journal} {\bibinfo
   {journal} {\apj}\ }\textbf {\bibinfo {volume} {705}},\ \bibinfo {pages}
  {659} (\bibinfo {year} {2009})},\ \Eprint {https://arxiv.org/abs/0810.3140}
  {arXiv:0810.3140 [astro-ph]} \BibitemShut {NoStop}%
\bibitem [{\citenamefont {{Oncins}}\ \emph {et~al.}(2022)\citenamefont
  {{Oncins}}, \citenamefont {{Miralda-Escud{\'e}}}, \citenamefont
  {{Guti{\'e}rrez}},\ and\ \citenamefont {{Gil-Pons}}}]{OncMGG22}%
  \BibitemOpen
  \bibfield  {author} {\bibinfo {author} {\bibfnamefont {M.}~\bibnamefont
  {{Oncins}}}, \bibinfo {author} {\bibfnamefont {J.}~\bibnamefont
  {{Miralda-Escud{\'e}}}}, \bibinfo {author} {\bibfnamefont {J.~L.}\
  \bibnamefont {{Guti{\'e}rrez}}},\ and\ \bibinfo {author} {\bibfnamefont
  {P.}~\bibnamefont {{Gil-Pons}}},\ }\bibfield  {title} {\bibinfo {title}
  {{Primordial black holes capture by stars and induced collapse to low-mass
  stellar black holes}},\ }\href {https://doi.org/10.1093/mnras/stac2647}
  {\bibfield  {journal} {\bibinfo  {journal} {\mnras}\ }\textbf {\bibinfo
  {volume} {517}},\ \bibinfo {pages} {28} (\bibinfo {year} {2022})},\ \Eprint
  {https://arxiv.org/abs/2205.13003} {arXiv:2205.13003 [astro-ph.GA]}
  \BibitemShut {NoStop}%
\bibitem [{\citenamefont {{Baumgarte}}\ and\ \citenamefont
  {{Shapiro}}(2025{\natexlab{a}})}]{BauS25c}%
  \BibitemOpen
  \bibfield  {author} {\bibinfo {author} {\bibfnamefont {T.~W.}\ \bibnamefont
  {{Baumgarte}}}\ and\ \bibinfo {author} {\bibfnamefont {S.~L.}\ \bibnamefont
  {{Shapiro}}},\ }\bibfield  {title} {\bibinfo {title} {{Can Premature Collapse
  Form Black Holes in the Upper and Lower Mass Gaps?}},\ }\href
  {https://doi.org/10.1103/26yd-1mhd} {\bibfield  {journal} {\bibinfo
  {journal} {\prl}\ }\textbf {\bibinfo {volume} {135}},\ \bibinfo {eid}
  {191401} (\bibinfo {year} {2025}{\natexlab{a}})},\ \Eprint
  {https://arxiv.org/abs/2509.04574} {arXiv:2509.04574 [astro-ph.HE]}
  \BibitemShut {NoStop}%
\bibitem [{\citenamefont {{McConnachie}}(2012)}]{McC12}%
  \BibitemOpen
  \bibfield  {author} {\bibinfo {author} {\bibfnamefont {A.~W.}\ \bibnamefont
  {{McConnachie}}},\ }\bibfield  {title} {\bibinfo {title} {{The Observed
  Properties of Dwarf Galaxies in and around the Local Group}},\ }\href
  {https://doi.org/10.1088/0004-6256/144/1/4} {\bibfield  {journal} {\bibinfo
  {journal} {\aj}\ }\textbf {\bibinfo {volume} {144}},\ \bibinfo {eid} {4}
  (\bibinfo {year} {2012})},\ \Eprint {https://arxiv.org/abs/1204.1562}
  {arXiv:1204.1562 [astro-ph.CO]} \BibitemShut {NoStop}%
\bibitem [{\citenamefont {{Simon}}(2019)}]{Sim19}%
  \BibitemOpen
  \bibfield  {author} {\bibinfo {author} {\bibfnamefont {J.~D.}\ \bibnamefont
  {{Simon}}},\ }\bibfield  {title} {\bibinfo {title} {{The Faintest Dwarf
  Galaxies}},\ }\href {https://doi.org/10.1146/annurev-astro-091918-104453}
  {\bibfield  {journal} {\bibinfo  {journal} {\araa}\ }\textbf {\bibinfo
  {volume} {57}},\ \bibinfo {pages} {375} (\bibinfo {year} {2019})},\ \Eprint
  {https://arxiv.org/abs/1901.05465} {arXiv:1901.05465 [astro-ph.GA]}
  \BibitemShut {NoStop}%
\bibitem [{\citenamefont {{Esser}}\ and\ \citenamefont
  {{Tinyakov}}(2023)}]{EssT23}%
  \BibitemOpen
  \bibfield  {author} {\bibinfo {author} {\bibfnamefont {N.}~\bibnamefont
  {{Esser}}}\ and\ \bibinfo {author} {\bibfnamefont {P.}~\bibnamefont
  {{Tinyakov}}},\ }\bibfield  {title} {\bibinfo {title} {{Constraints on
  primordial black holes from observation of stars in dwarf galaxies}},\ }\href
  {https://doi.org/10.1103/PhysRevD.107.103052} {\bibfield  {journal} {\bibinfo
   {journal} {\prd}\ }\textbf {\bibinfo {volume} {107}},\ \bibinfo {eid}
  {103052} (\bibinfo {year} {2023})},\ \Eprint
  {https://arxiv.org/abs/2207.07412} {arXiv:2207.07412 [astro-ph.HE]}
  \BibitemShut {NoStop}%
\bibitem [{\citenamefont {{Walker}}\ \emph {et~al.}(2007)\citenamefont
  {{Walker}}, \citenamefont {{Mateo}}, \citenamefont {{Olszewski}},
  \citenamefont {{Gnedin}}, \citenamefont {{Wang}}, \citenamefont {{Sen}},\
  and\ \citenamefont {{Woodroofe}}}]{WalMOGWSW07}%
  \BibitemOpen
  \bibfield  {author} {\bibinfo {author} {\bibfnamefont {M.~G.}\ \bibnamefont
  {{Walker}}}, \bibinfo {author} {\bibfnamefont {M.}~\bibnamefont {{Mateo}}},
  \bibinfo {author} {\bibfnamefont {E.~W.}\ \bibnamefont {{Olszewski}}},
  \bibinfo {author} {\bibfnamefont {O.~Y.}\ \bibnamefont {{Gnedin}}}, \bibinfo
  {author} {\bibfnamefont {X.}~\bibnamefont {{Wang}}}, \bibinfo {author}
  {\bibfnamefont {B.}~\bibnamefont {{Sen}}},\ and\ \bibinfo {author}
  {\bibfnamefont {M.}~\bibnamefont {{Woodroofe}}},\ }\bibfield  {title}
  {\bibinfo {title} {{Velocity Dispersion Profiles of Seven Dwarf Spheroidal
  Galaxies}},\ }\href {https://doi.org/10.1086/521998} {\bibfield  {journal}
  {\bibinfo  {journal} {\apjl}\ }\textbf {\bibinfo {volume} {667}},\ \bibinfo
  {pages} {L53} (\bibinfo {year} {2007})},\ \Eprint
  {https://arxiv.org/abs/0708.0010} {arXiv:0708.0010 [astro-ph]} \BibitemShut
  {NoStop}%
\bibitem [{\citenamefont {{Battaglia}}\ \emph {et~al.}(2013)\citenamefont
  {{Battaglia}}, \citenamefont {{Helmi}},\ and\ \citenamefont
  {{Breddels}}}]{BatHB13}%
  \BibitemOpen
  \bibfield  {author} {\bibinfo {author} {\bibfnamefont {G.}~\bibnamefont
  {{Battaglia}}}, \bibinfo {author} {\bibfnamefont {A.}~\bibnamefont
  {{Helmi}}},\ and\ \bibinfo {author} {\bibfnamefont {M.}~\bibnamefont
  {{Breddels}}},\ }\bibfield  {title} {\bibinfo {title} {{Internal kinematics
  and dynamical models of dwarf spheroidal galaxies around the Milky Way}},\
  }\href {https://doi.org/10.1016/j.newar.2013.05.003} {\bibfield  {journal}
  {\bibinfo  {journal} {\nar}\ }\textbf {\bibinfo {volume} {57}},\ \bibinfo
  {pages} {52} (\bibinfo {year} {2013})},\ \Eprint
  {https://arxiv.org/abs/1305.5965} {arXiv:1305.5965 [astro-ph.CO]}
  \BibitemShut {NoStop}%
\bibitem [{\citenamefont {{Strigari}}\ \emph {et~al.}(2008)\citenamefont
  {{Strigari}}, \citenamefont {{Bullock}}, \citenamefont {{Kaplinghat}},
  \citenamefont {{Simon}}, \citenamefont {{Geha}}, \citenamefont {{Willman}},\
  and\ \citenamefont {{Walker}}}]{Stretal08}%
  \BibitemOpen
  \bibfield  {author} {\bibinfo {author} {\bibfnamefont {L.~E.}\ \bibnamefont
  {{Strigari}}}, \bibinfo {author} {\bibfnamefont {J.~S.}\ \bibnamefont
  {{Bullock}}}, \bibinfo {author} {\bibfnamefont {M.}~\bibnamefont
  {{Kaplinghat}}}, \bibinfo {author} {\bibfnamefont {J.~D.}\ \bibnamefont
  {{Simon}}}, \bibinfo {author} {\bibfnamefont {M.}~\bibnamefont {{Geha}}},
  \bibinfo {author} {\bibfnamefont {B.}~\bibnamefont {{Willman}}},\ and\
  \bibinfo {author} {\bibfnamefont {M.~G.}\ \bibnamefont {{Walker}}},\
  }\bibfield  {title} {\bibinfo {title} {{A common mass scale for satellite
  galaxies of the Milky Way}},\ }\href {https://doi.org/10.1038/nature07222}
  {\bibfield  {journal} {\bibinfo  {journal} {\nat}\ }\textbf {\bibinfo
  {volume} {454}},\ \bibinfo {pages} {1096} (\bibinfo {year} {2008})},\ \Eprint
  {https://arxiv.org/abs/0808.3772} {arXiv:0808.3772 [astro-ph]} \BibitemShut
  {NoStop}%
\bibitem [{\citenamefont {{Read}}\ \emph {et~al.}(2019)\citenamefont {{Read}},
  \citenamefont {{Walker}},\ and\ \citenamefont {{Steger}}}]{ReaWS19}%
  \BibitemOpen
  \bibfield  {author} {\bibinfo {author} {\bibfnamefont {J.~I.}\ \bibnamefont
  {{Read}}}, \bibinfo {author} {\bibfnamefont {M.~G.}\ \bibnamefont
  {{Walker}}},\ and\ \bibinfo {author} {\bibfnamefont {P.}~\bibnamefont
  {{Steger}}},\ }\bibfield  {title} {\bibinfo {title} {{Dark matter heats up in
  dwarf galaxies}},\ }\href {https://doi.org/10.1093/mnras/sty3404} {\bibfield
  {journal} {\bibinfo  {journal} {\mnras}\ }\textbf {\bibinfo {volume} {484}},\
  \bibinfo {pages} {1401} (\bibinfo {year} {2019})},\ \Eprint
  {https://arxiv.org/abs/1808.06634} {arXiv:1808.06634 [astro-ph.GA]}
  \BibitemShut {NoStop}%
\bibitem [{\citenamefont {{Baumgarte}}\ and\ \citenamefont
  {{Shapiro}}(2024{\natexlab{a}})}]{BauS24d}%
  \BibitemOpen
  \bibfield  {author} {\bibinfo {author} {\bibfnamefont {T.~W.}\ \bibnamefont
  {{Baumgarte}}}\ and\ \bibinfo {author} {\bibfnamefont {S.~L.}\ \bibnamefont
  {{Shapiro}}},\ }\bibfield  {title} {\bibinfo {title} {{Primordial black holes
  captured by neutron stars: Simulations in general relativity}},\ }\href
  {https://doi.org/10.1103/PhysRevD.110.023021} {\bibfield  {journal} {\bibinfo
   {journal} {\prd}\ }\textbf {\bibinfo {volume} {110}},\ \bibinfo {eid}
  {023021} (\bibinfo {year} {2024}{\natexlab{a}})},\ \Eprint
  {https://arxiv.org/abs/2405.10365} {arXiv:2405.10365 [gr-qc]} \BibitemShut
  {NoStop}%
\bibitem [{\citenamefont {{Baumgarte}}\ and\ \citenamefont
  {{Shapiro}}(2025{\natexlab{b}})}]{BauS25a}%
  \BibitemOpen
  \bibfield  {author} {\bibinfo {author} {\bibfnamefont {T.~W.}\ \bibnamefont
  {{Baumgarte}}}\ and\ \bibinfo {author} {\bibfnamefont {S.~L.}\ \bibnamefont
  {{Shapiro}}},\ }\bibfield  {title} {\bibinfo {title} {{Boosting the growth of
  intermediate-mass black holes: Collisions with massive stars}},\ }\href
  {https://doi.org/10.1103/PhysRevD.111.063039} {\bibfield  {journal} {\bibinfo
   {journal} {\prd}\ }\textbf {\bibinfo {volume} {111}},\ \bibinfo {eid}
  {063039} (\bibinfo {year} {2025}{\natexlab{b}})},\ \Eprint
  {https://arxiv.org/abs/2502.14955} {arXiv:2502.14955 [astro-ph.HE]}
  \BibitemShut {NoStop}%
\bibitem [{\citenamefont {{Abramowicz}}\ \emph {et~al.}(2018)\citenamefont
  {{Abramowicz}}, \citenamefont {{Bejger}},\ and\ \citenamefont
  {{Wielgus}}}]{AbrBW18}%
  \BibitemOpen
  \bibfield  {author} {\bibinfo {author} {\bibfnamefont {M.~A.}\ \bibnamefont
  {{Abramowicz}}}, \bibinfo {author} {\bibfnamefont {M.}~\bibnamefont
  {{Bejger}}},\ and\ \bibinfo {author} {\bibfnamefont {M.}~\bibnamefont
  {{Wielgus}}},\ }\bibfield  {title} {\bibinfo {title} {{Collisions of Neutron
  Stars with Primordial Black Holes as Fast Radio Bursts Engines}},\ }\href
  {https://doi.org/10.3847/1538-4357/aae64a} {\bibfield  {journal} {\bibinfo
  {journal} {\apj}\ }\textbf {\bibinfo {volume} {868}},\ \bibinfo {eid} {17}
  (\bibinfo {year} {2018})},\ \Eprint {https://arxiv.org/abs/1704.05931}
  {arXiv:1704.05931 [astro-ph.HE]} \BibitemShut {NoStop}%
\bibitem [{\citenamefont {{Baumgarte}}\ and\ \citenamefont
  {{Shapiro}}(2024{\natexlab{b}})}]{BauS24c}%
  \BibitemOpen
  \bibfield  {author} {\bibinfo {author} {\bibfnamefont {T.~W.}\ \bibnamefont
  {{Baumgarte}}}\ and\ \bibinfo {author} {\bibfnamefont {S.~L.}\ \bibnamefont
  {{Shapiro}}},\ }\bibfield  {title} {\bibinfo {title} {{Primordial black holes
  captured by neutron stars: Relativistic point-mass treatment}},\ }\href
  {https://doi.org/10.1103/PhysRevD.109.123012} {\bibfield  {journal} {\bibinfo
   {journal} {\prd}\ }\textbf {\bibinfo {volume} {109}},\ \bibinfo {eid}
  {123012} (\bibinfo {year} {2024}{\natexlab{b}})},\ \Eprint
  {https://arxiv.org/abs/2404.08735} {arXiv:2404.08735 [gr-qc]} \BibitemShut
  {NoStop}%
\bibitem [{\citenamefont {{Bondi}}(1952)}]{Bon52}%
  \BibitemOpen
  \bibfield  {author} {\bibinfo {author} {\bibfnamefont {H.}~\bibnamefont
  {{Bondi}}},\ }\bibfield  {title} {\bibinfo {title} {{On spherically
  symmetrical accretion}},\ }\href {https://doi.org/10.1093/mnras/112.2.195}
  {\bibfield  {journal} {\bibinfo  {journal} {\mnras}\ }\textbf {\bibinfo
  {volume} {112}},\ \bibinfo {pages} {195} (\bibinfo {year}
  {1952})}\BibitemShut {NoStop}%
\bibitem [{\citenamefont {{Shapiro}}\ and\ \citenamefont
  {{Teukolsky}}(1983)}]{ShaT83}%
  \BibitemOpen
  \bibfield  {author} {\bibinfo {author} {\bibfnamefont {S.~L.}\ \bibnamefont
  {{Shapiro}}}\ and\ \bibinfo {author} {\bibfnamefont {S.~A.}\ \bibnamefont
  {{Teukolsky}}},\ }\href {https://doi.org/10.1002/9783527617661} {\emph
  {\bibinfo {title} {{Black holes, white dwarfs and neutron stars. The physics
  of compact objects}}}}\ (\bibinfo  {publisher} {Wiley Interscience},\
  \bibinfo {year} {1983})\BibitemShut {NoStop}%
\bibitem [{\citenamefont {{Richards}}\ \emph {et~al.}(2021)\citenamefont
  {{Richards}}, \citenamefont {{Baumgarte}},\ and\ \citenamefont
  {{Shapiro}}}]{RicBS21}%
  \BibitemOpen
  \bibfield  {author} {\bibinfo {author} {\bibfnamefont {C.~B.}\ \bibnamefont
  {{Richards}}}, \bibinfo {author} {\bibfnamefont {T.~W.}\ \bibnamefont
  {{Baumgarte}}},\ and\ \bibinfo {author} {\bibfnamefont {S.~L.}\ \bibnamefont
  {{Shapiro}}},\ }\bibfield  {title} {\bibinfo {title} {{Accretion onto a small
  black hole at the center of a neutron star}},\ }\href
  {https://doi.org/10.1103/PhysRevD.103.104009} {\bibfield  {journal} {\bibinfo
   {journal} {\prd}\ }\textbf {\bibinfo {volume} {103}},\ \bibinfo {eid}
  {104009} (\bibinfo {year} {2021})},\ \Eprint
  {https://arxiv.org/abs/2102.09574} {arXiv:2102.09574 [astro-ph.HE]}
  \BibitemShut {NoStop}%
\bibitem [{\citenamefont {{Chandrasekhar}}(1967)}]{Cha67}%
  \BibitemOpen
  \bibfield  {author} {\bibinfo {author} {\bibfnamefont {S.}~\bibnamefont
  {{Chandrasekhar}}},\ }\href@noop {} {\emph {\bibinfo {title} {{An
  introduction to the study of stellar structure}}}}\ (\bibinfo  {publisher}
  {Dover},\ \bibinfo {year} {1967})\BibitemShut {NoStop}%
\bibitem [{\citenamefont {{East}}\ and\ \citenamefont
  {{Lehner}}(2019)}]{EasL19}%
  \BibitemOpen
  \bibfield  {author} {\bibinfo {author} {\bibfnamefont {W.~E.}\ \bibnamefont
  {{East}}}\ and\ \bibinfo {author} {\bibfnamefont {L.}~\bibnamefont
  {{Lehner}}},\ }\bibfield  {title} {\bibinfo {title} {{Fate of a neutron star
  with an endoparasitic black hole and implications for dark matter}},\ }\href
  {https://doi.org/10.1103/PhysRevD.100.124026} {\bibfield  {journal} {\bibinfo
   {journal} {\prd}\ }\textbf {\bibinfo {volume} {100}},\ \bibinfo {eid}
  {124026} (\bibinfo {year} {2019})},\ \Eprint
  {https://arxiv.org/abs/1909.07968} {arXiv:1909.07968 [gr-qc]} \BibitemShut
  {NoStop}%
\bibitem [{\citenamefont {{Esser}}\ \emph {et~al.}(2024)\citenamefont
  {{Esser}}, \citenamefont {{De Rijcke}},\ and\ \citenamefont
  {{Tinyakov}}}]{EssDT24}%
  \BibitemOpen
  \bibfield  {author} {\bibinfo {author} {\bibfnamefont {N.}~\bibnamefont
  {{Esser}}}, \bibinfo {author} {\bibfnamefont {S.}~\bibnamefont {{De
  Rijcke}}},\ and\ \bibinfo {author} {\bibfnamefont {P.}~\bibnamefont
  {{Tinyakov}}},\ }\bibfield  {title} {\bibinfo {title} {{The impact of
  primordial black holes on the stellar mass function of ultra-faint dwarf
  galaxies}},\ }\href {https://doi.org/10.1093/mnras/stae147} {\bibfield
  {journal} {\bibinfo  {journal} {\mnras}\ }\textbf {\bibinfo {volume} {529}},\
  \bibinfo {pages} {32} (\bibinfo {year} {2024})},\ \Eprint
  {https://arxiv.org/abs/2311.12658} {arXiv:2311.12658 [astro-ph.GA]}
  \BibitemShut {NoStop}%
\end{thebibliography}


\end{document}